# Semantic optical fiber communication system

**Author list:** Zhenming Yu[1,2,*], Hongyu Huang[1,2], Liming Cheng[1,2], Wei Zhang[1], Yueqiu Mu[1], and Kun Xu[1,*]

\* Corresponding author: yuzhenming@bupt.edu.cn; xukun@bupt.edu.cn

**Affiliations:**

[1]State Key Laboratory of Information Photonics and Optical Communications, Beijing University of Posts and Telecommunications, Beijing 100876, China

[2]These authors contributed equally: Zhenming Yu, Hongyu Huang, and Liming Cheng.

## Abstract

The current optical communication systems minimize bit or symbol errors without considering the semantic meaning behind digital bits, thus transmitting a lot of unnecessary information. We propose and experimentally demonstrate a semantic optical fiber communication (SOFC) system. Instead of encoding information into bits for transmission, semantic information is extracted from the source using deep learning. The generated semantic symbols are then directly transmitted through an optical fiber. Compared with the bit-based structure, the SOFC system achieved higher information compression and a more stable performance, especially in the low received optical power regime, and enhanced the robustness against optical link impairments. This work introduces an intelligent optical communication system at the human analytical thinking level, which is a significant step toward a breakthrough in the current optical communication architecture.

## Introduction

Optical fiber communication is the cornerstone of network information transmission, carrying more than 90% of the world's data traffic [1, 2]. Owing to the increasing demand for higher transmission capacity, advanced coding, modulation, multiplexing, receiver, and compensation techniques have been introduced [3-15], promoting the capacity of optical communication systems to approach the Shannon limit [16]. However, improving capacity with limited spectrum and power consumption has hindered the ongoing evolution. As the global internet traffic increases by 60% per year [17], theoretical breakthroughs in optical fiber communication are imperative to meet future demands.

As summarized by Shannon and Weaver, communication could be categorized into three levels [18, 19]:

- First level: How accurately can communication symbols be transmitted? (The technical problem)

- Second level: How precisely do transmitted symbols convey the desired meaning? (The semantic problem)
- Third level: How effectively does the received meaning affect conduct in the desired way? (The effectiveness problem)

Until now, optical fiber communication systems are still at the first level, that is, encoding the information source into bits, modulating bits into symbols, and guaranteeing the transmitted symbols as accurately as possible. In this structure, the meanings behind the transmitted bits are almost irrelevant to communication, causing the transmission of a lot of redundant information, which cannot best fit the information source and optical physical channel.

Semantic information is considered "the meaningful messages characterizing the observations of the world" [20]. Efficient transmission of semantic information is then on the point of upgrading to the second level of communication [21, 22]. The most notable feature of semantic communication is to draw inspiration from human language communication and focus on delivering the meaning of messages. The goal of semantic transmission system is no longer error-free transmission but delivering the primary semantic information that contributes to human understanding [23-25]. To break the development bottleneck in optical fiber communication, it is urgent to study SOFC systems, which have great potential to further increase the capacity and transmission distance, and contribute to the realization of the more intelligent and briefer optical network.

Here, we propose and experimentally demonstrate a novel framework of SOFC system. The semantic transmission of two information sources, namely texts and images, through the intensity modulation/direct detection (IM/DD)-based optical fiber link is realized. In text transmission, the language attention network (LA-net) is designed to restore the meaning of sentences and minimize semantic errors. In image transmission, the dual-attention residual network (DR-net) is designed to extract rich semantic features from images and minimize semantic errors. To make the semantic decoding robust against optical link impairments, we incorporated a convolutional neural network (CNN) into the semantic decoding network and implemented joint optimization (JO). For comparison, we conducted experiments on the traditional IM/DD PAM8 and PAM4 optical fiber communication systems. The results showed that the SOFC system achieved higher information compression and was more robust to Gaussian noise and optical link impairments. Especially when the optical channel environment was harsh, the performance of the traditional optical fiber communication (TOFC) systems declined by a "cliff," while that of the SOFC system was stable.

**Structure**

The semantic encoder and decoder are trained by end to end learning firstly (Fig. 1a). White Gaussian noise is added to output symbols of the encoder. To further enhance the robustness, the noise intensity is randomly changed, conforming with the uniform distribution within the SNR range of 4 to 20 dB. The generated semantic symbols are sent to the optical fiber for transmission directly. The SOFC system was established based on the IM-DD optical fiber transmission system.

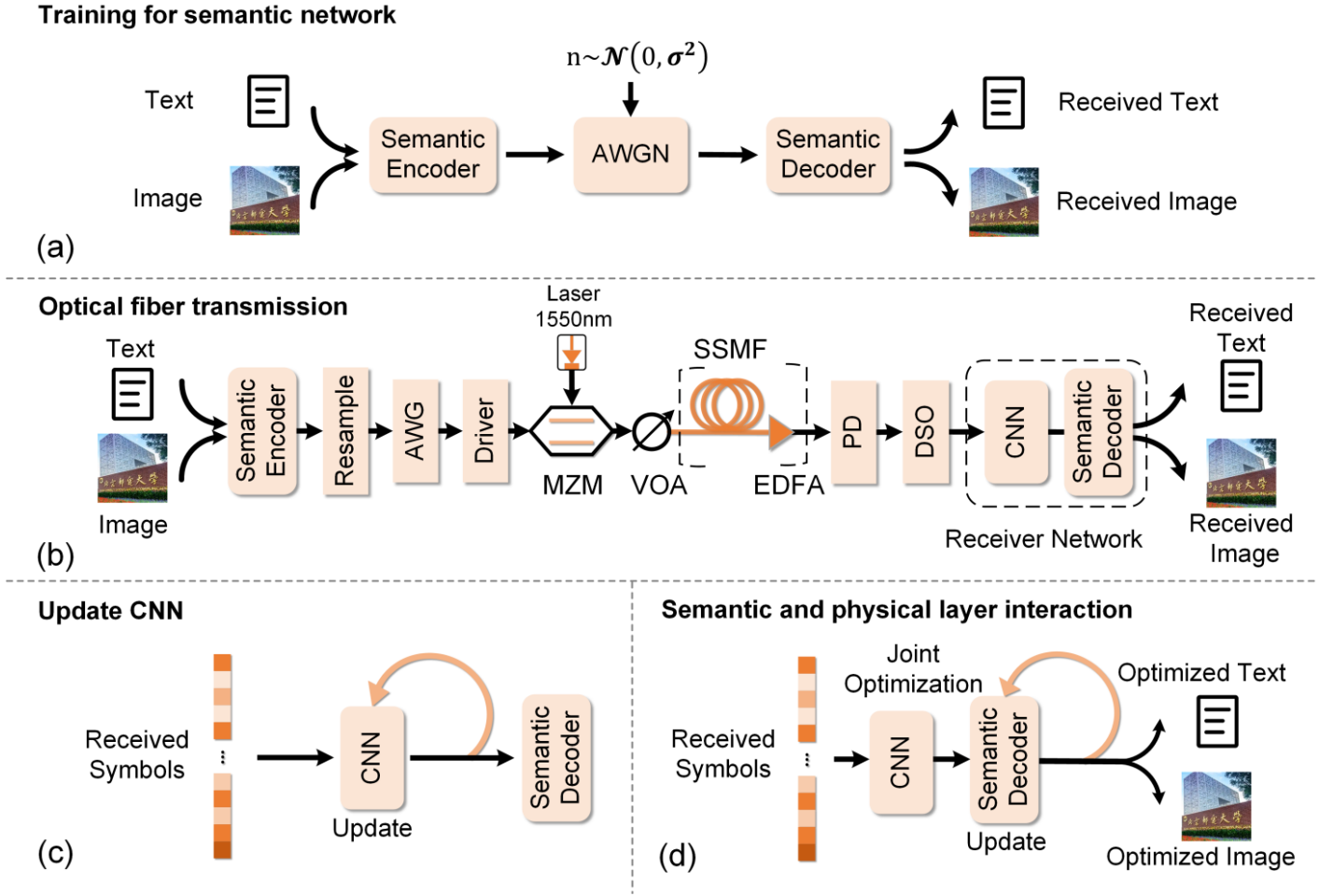

**Fig. 1** (a) Learning phase of the semantic encoder and decoder under the AWGN channel. (b) Experimental setup for optical transmission. (c) Parameter updating of the CNN equalization algorithm. (d) Parameter updating of the semantic decoder for joint optimization. AWG: arbitrary waveform generator; MZM: Mach-Zehnder modulator; SSMF: standard single-mode fiber; VOA: variable optical attenuator; EDFA: erbium-doped fiber amplifier; PD: photodetector; DSO: digital sampling oscilloscope; CNN: convolutional neural network.

The experimental setup of the optical transmission system is shown in Fig. 1(b). At the transmitter, the output symbols of the semantic encoder are processed by resampling and a square-root-raised cosine. In the presence of a bandwidth limitation (BWL), digital pre-emphasis is adopted. The signals are then sent to the arbitrary waveform generator (AWG, Keysight M8195A). The highest sampling rate and analog bandwidth of this AWG are 65 GSa/s and 25 GHz, respectively. The electrical analog signals generated by the AWG are amplified by a linear broadband amplifier (SHF S807C). The electrical signals are then modulated into optical signals by a Mach-Zehnder modulator (EOSPACE AX-0MVS-40-PFA-PFA-LV). In B2B transmission, a variable optical attenuator (VOA) is utilized to control the ROP. When evaluating the tolerance to peak-to-peak values and baud rates of RF signals, the ROP was set as 0 and −4 dBm, respectively. In fiber transmission, the input fiber optical power and ROP were both 0 dBm, with the control of the VOA and erbium-doped fiber amplifier (EDFA). At the receiver, the optical signals were converted into electrical signals by the photodetector (PD) with a 30-GHz bandwidth (FINISAR XPRV2325A). The received electrical signals were sampled at 100 GSa/s using the digital sampling oscilloscope (DSO) with an analog bandwidth of 25 GHz (Tektronix DSA72504D). Then the digital signals were processed by the receiver network.

To alleviate the distortions in transmission, a CNN is incorporated into the semantic decoding network to recover the information source jointly. The proposed CNN is composed of one input layer and three convolutional layers, whose parameters are updated by supervised training using part of the received symbols (Fig. 1c). To further improve the performance, JO is initiated; that is, optical physical layer equalization and semantic decoding are integrated (Fig. 1d). Specifically, after processing with the trained CNN, the semantic decoding network parameters are updated again. This operation enables the interaction between the optical physical channel and the semantic information, which could promote the adaptability of the semantic decoder to the real optical channel environment. The semantic symbols are then fed into the updated semantic decoder for semantic source reconstruction.

**Text transmission**

In the SOFC system, errors can be accepted without affecting human understanding. For example, if the sentence, "My car is parked in the school," is transmitted, the received sentence could be "My automobile that parked in the school" or "My car is parked in a school". Word errors do not affect the understanding of the sentence. By sacrificing information unessential to human understanding, the efficiency of communication can be improved further.

We designed the LA-net as the semantic codec of text. The transformer network is used as the backbone, which is popular in natural language processing (NLP) [26]. The core of the transformer network is the multi-head self-attention mechanism, which enables the effective learning of semantic information [27]. As shown in Fig. 2(a), the attention mechanism learns that "it" represents "the cat". During sentence reconstruction, the attention mechanism could determine that the word after "because" has similar semantic information to "the cat". This operation will contribute to the prediction of transmitted words.

The designed structures of the semantic encoder and decoder are shown in Fig. 2(b). The encoding process is as follows: First, the text is segmented by word. Then, the words are converted into one hot vector. After that, the one hot vector is encoded by the semantic encoder. The embedding layer in the encoder can reduce the dimension of the input one hot vector. Three transformers are utilized to extract the semantic information. Two full-connected neural networks (FCNNs) act as an auto-encoder, further compressing the information. The activation function of the last layer in the encoder is designed as HardTanh, which limits the amplitude of the output signal to [−1,1] and avoid excessive signal peak-to-average power ratio (PAPR) [28]. The decoding process is the inverse of encoding.

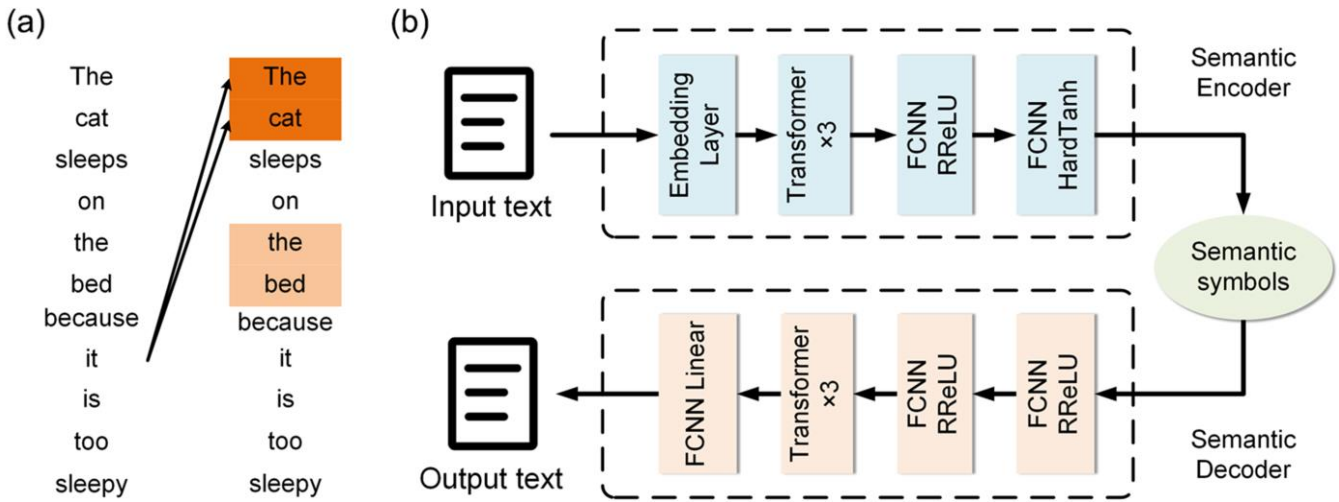

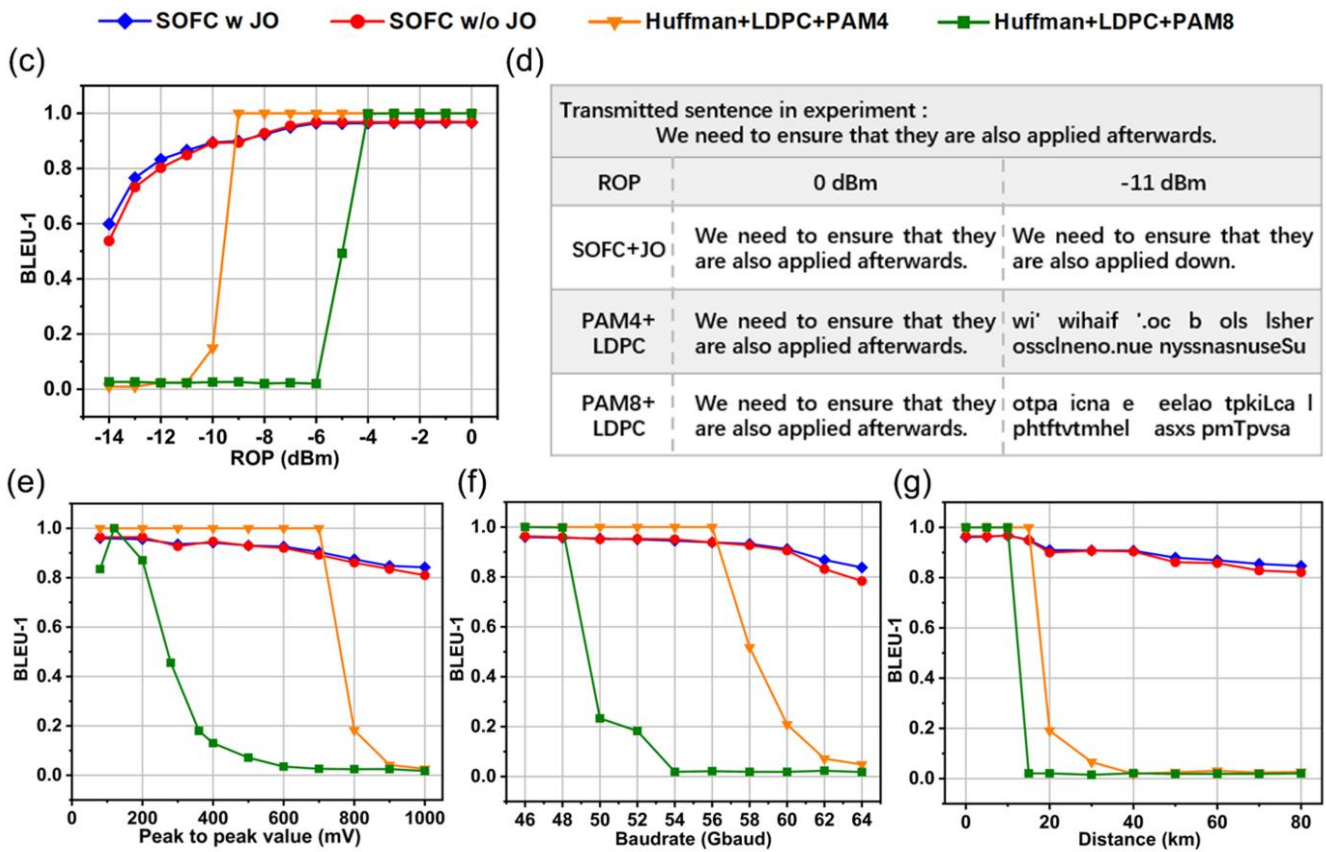

**Fig. 2 (**a) Example of the multi-head self-attention mechanism. (b) Network structure of the semantic encoder and decoder. (c) Experimental results with respect to the ROP. (d) Sample sentences with ROPs of 0 and −11 dBm. Transmission performance with respect to (e) peak-to-peak value, (f) baud rate, and (g) transmission distance. FCNN: full-connected neural network; ROP: received optical power.

The bit-based TOFC system was used for comparison. PAM4 and PAM8 were used as the modulation formats. Texts were encoded into bits by Huffman coding [29]. The 2/3 low-density parity check (LDPC) was utilized for channel coding. The 61-tap feed-forward equalization (FFE) was used for channel equalization.

The symbol and sampling rates were 30 Gbaud and 60 GSa/s, respectively. Therefore, the effective bit rate of PAM8 transmission is 60 Gbit/s, which can deliver approximately $1.25 \times 10^8$ sentences/s (each sentence within 30 words). The effective bit rate of PAM4 transmission is 40 Gbit/s, which can deliver approximately $8.33 \times 10^7$ sentences/s. In the SOFC system, the effective data rate is approximately $1.25 \times 10^8$ sentences/s, equivalent to that of PAM8 transmission.

The text transmission performance was evaluated on the basis of the bilingual evaluation understudy (BLEU) score [30]. The BLEU under different received optical power (ROP), baud rate, and peak-to-peak value of radio frequency (RF) signals were obtained in optical back-to-back (B2B) transmission. The relation between the calculated BLEU score and ROP is shown in Fig. 2(c). The results indicate the SOFC system is more robust to noise. With JO, the SOFC system performance can be improved further. When the ROP was below −5 dBm, the PAM8 transmission performance declined by a "cliff." However, the SOFC method provides a stable performance degradation. When the ROP was higher than −11 dBm, the BLEU-1 score was higher than 0.85, which indicates that the word error rate (WER) was less than 0.15. By setting a BLEU-1 score of 0.85 as the standard, the receiver sensitivity was increased by approximately 6 dB in the SOFC system compared with the TOFC system at the same rate and by approximately 2 dB at 1.5 times faster rate. Fig. 2(d) shows the transmission result of a sample sentence. When the ROP was −11 dBm, the recovered sentence was completely garbled in the TOFC systems, whereas the sentence meaning was recovered well in the SOFC system despite word errors.

Fig. 2(e) shows the tolerance to modulation nonlinearity. With the increasing peak-to-peak value of the modulated signals, the transfer function of the optical modulator will enter the nonlinear region. The results indicate the SOFC system performance is less affected by the modulation nonlinear noise. Fig. 2(f) shows the tolerance to the bandwidth limit. When the AWG and DSO both have a 25-GHz analog bandwidth, the SOFC system enabled the transmission of 64 Gbaud, while the PAM4 and PAM8 transmissions achieved up to 56 and 48 Gbaud, respectively. Fig. 2(g) shows the tolerance to fiber transmission distance. As the transmission distance increases, the accumulated chromatic dispersion (CD) will degrade the signal quality seriously [31]. Without special CD compensation, reliable PAM4 and PAM8 transmissions are only possible within 20 km. By contrast, the SOFC system can transmit the text over 80 km without losing the main semantic information.

**Image transmission**

Natural images contain abundant semantic information according to human comprehension, as depicted in Fig. 3(a), where the basic semantic components reveal the visual structure and holistic relation of an image.

In the SOFC system, the scheme that delivering the underlying meaning of images is established. We adopted deep CNN for semantic coding, inspired by its impressive performance in computer versions (CV) [32].

We designed the DR-net to extract semantics from image sources. Its overall structure is shown in Fig. 3(b). Residual convolution blocks [33] and the attention mechanism were implemented for semantic extraction. By sliding the convolution kernel over the image data, a feature map can be generated. Suppose the input image is $I \in \mathbb{R}^{H \times W \times 3}$, after passing through three convolution blocks, semantic tokens $F \in \mathbb{R}^{H/4 \times W/4 \times 128}$ are obtained. By flattening $F$ into a one-dimensional vector $F^* \in \mathbb{R}^N$ ($N = H/4 \times W/4 \times 128$) and applying a fully connected layer, the semantic signal $S \in \mathbb{R}^L$ is generated. The decoding process goes with the inverse operation. Fig. 3(c) shows the semantics flow of the encoding phase. By transforming the input image into a high-level semantic space, abundant semantic tokens were generated, carrying significant structures and context information of the encoded image. The extracted semantic tokens were then mapped into semantic symbols before they were sent to a real optical physical channel. The semantic decoder was designed with a symmetrical structure. Two types of attention mechanisms are implemented in the DR-net. The dual-attention [34] module tends to guide the network to focus on the effective semantics underlying the input image, while the SNR attention [35] module leads the network to protect the significant semantics against optical channel noise.

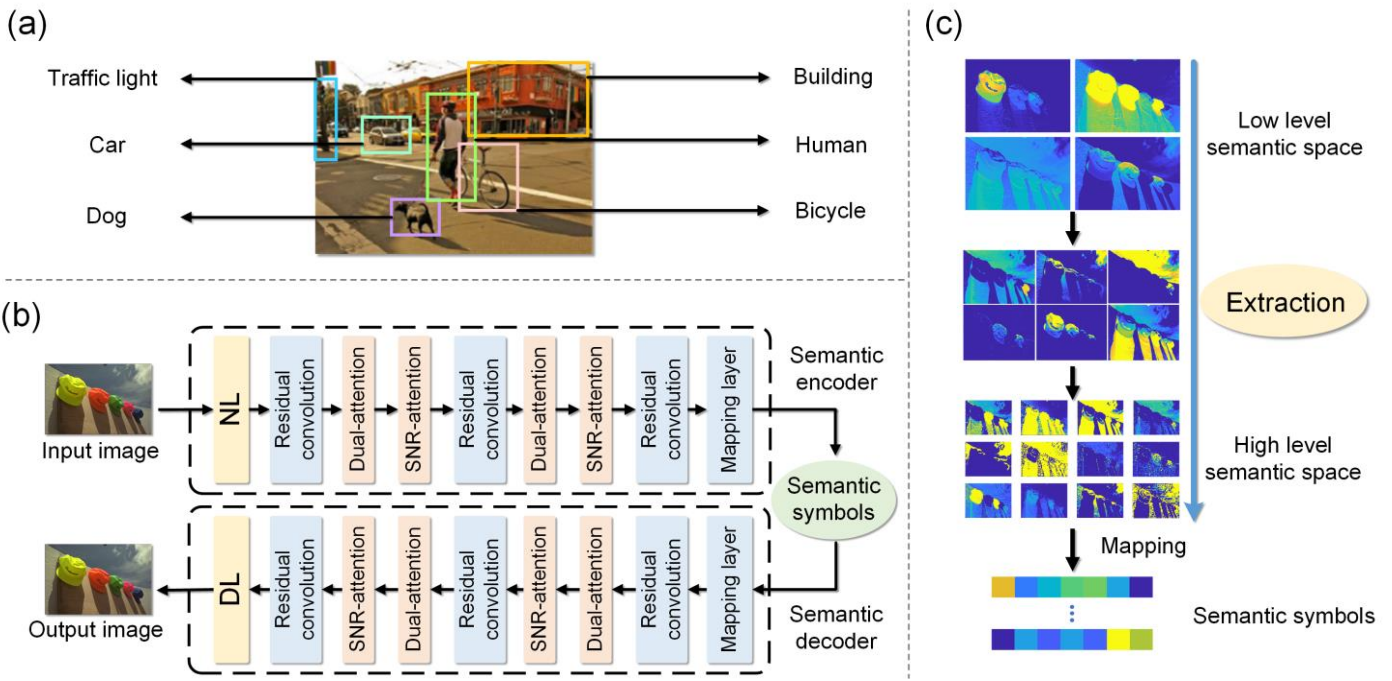

**Fig. 3** (a) Semantics of an image based on human comprehension. (b) Overview of the designed DR-net. (c) Image semantics flow. NL: normalization layer; DL: denormalization layer; Res-block: residual convolution block.

For implementation of the TOFC schemes, we selected the joint photographic experts group (JPEG) [36] and JPEG2000 [37] (JP2K) for source coding. The same channel coding, modulation formats, channel equalization method, and experimental setup were adopted, as in the text transmission task.

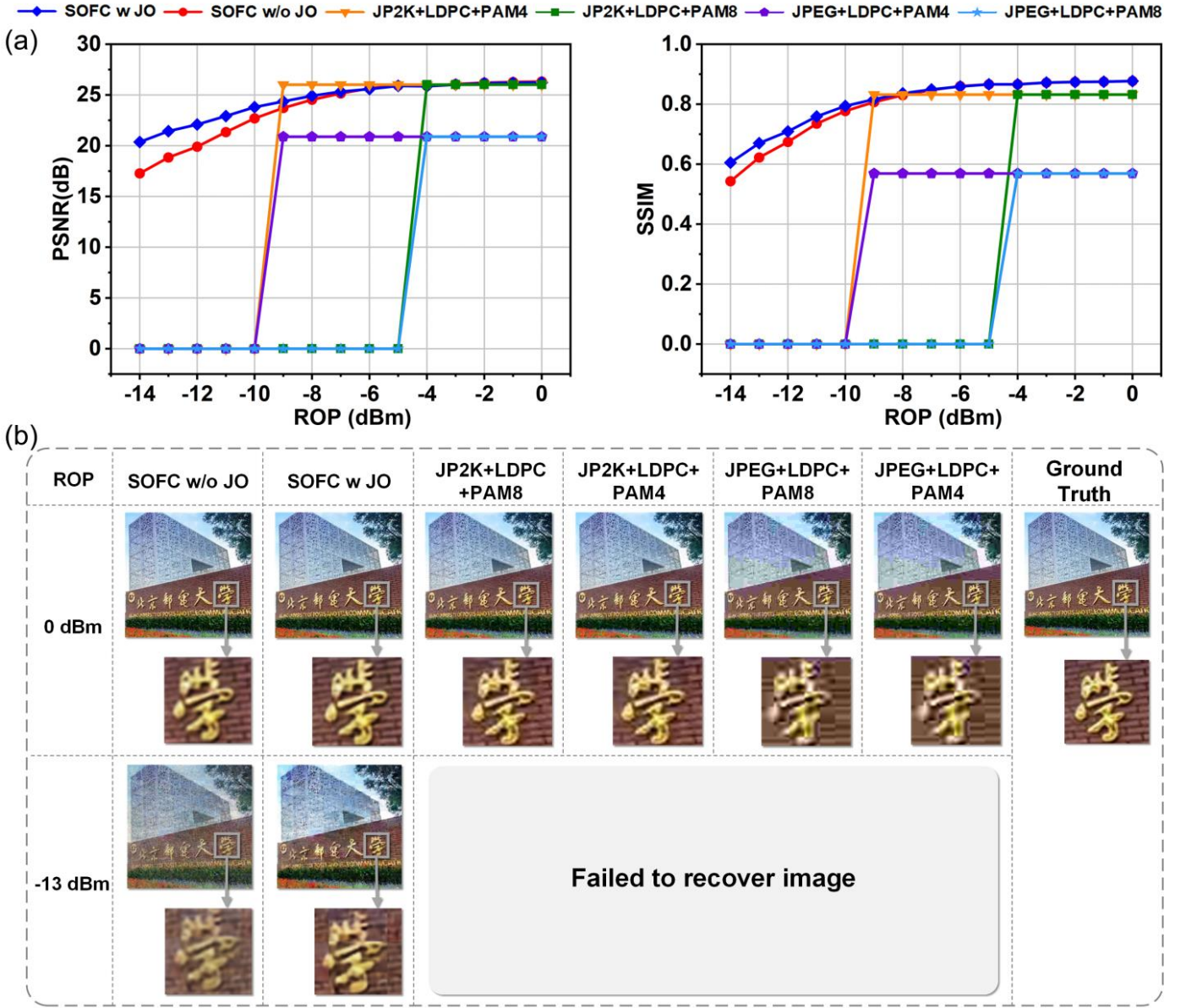

**Fig. 4** (a) Experimental results with respect to the ROP. (b) Decoded images when the ROPs were 0 and −13 dBm.

A real-world RGB image ($256{\times}256{\times}3$; Fig. 4(b)) was used for transmission. Through the semantic encoder, the image was encoded into a semantic symbol sequence with a length of 34,368, and the value of each symbol is normalized to $[-1, 1]$. The data rate of this SOFC system is $8.73{\times}10^5$ images/s. As for the traditional methods, when PAM8 was applied, we adjusted the compression ratio of JPEG and JP2K to keep the data rate equivalent to that of the SOFC system, which is $8.73{\times}10^5$ images/s. When PAM4 was applied, the data rate was a third lower at $5.82{\times}10^5$ images/s.

To quantitatively evaluate the transmission quality, the peak signal-to-noise ratio (PSNR) and structural similarity index matrix (SSIM) [38] were calculated. The transmission results under different ROPs, baud rates, and peak-to-peak values of RF signals were obtained in optical B2B transmission. Fig. 4(a) shows the PSNR and SSIM according to the ROPs. It illustrates that the SOFC system is more robust to channel noise, and its performance can be further improved with JO. When evaluated using SSIM, the superiority of the

SOFC system is proved more effectively because the SSIM criterion focuses on the semantic structure of an image. The "cliff" effect occurs when the ROP is −5 dBm for PAM8 and −10 dBm for PAM4, while the acceptable performance can still be achieved with the SOFC system. By setting a 20-dB PSNR as the standard, the receiver sensitivity was increased by approximately 10 dB in the SOFC system compared with the TOFC at the same rate and by approximately 5 dB at 1.5 times faster rate (Fig. 4(a)). Fig. 4(b) shows the decoded images at the receiver with ROPs of 0 and −13 dBm. When the ROP was 0 dBm, the SOFC system provided a high transmission fidelity, whereas the traditional schemes showed different types of degradation. When the ROP was −13 dBm, the image could not be decoded at all in the TOFC systems. A PSNR greater than 20 dB and a SSIM greater than 0.7 were still achieved in the SOFC system.

Fig. 5(a) shows the tolerance to modulation nonlinearity, which indicates that the SOFC system is more robust to the nonlinear noise of modulation. Fig. 5(b) shows the tolerance to the bandwidth limit. The SOFC system enabled the transmission of 64 Gbaud, whereas the PAM4 and PAM8 schemes achieved up to 56 and 46 Gbaud, respectively. Fig. 5(c) shows the tolerance to transmission distance. The PAM4 and PAM8 schemes can reliably transmit the image within 20 km, whereas the SOFC system can transmit the same image over 80 km without severe degradation.

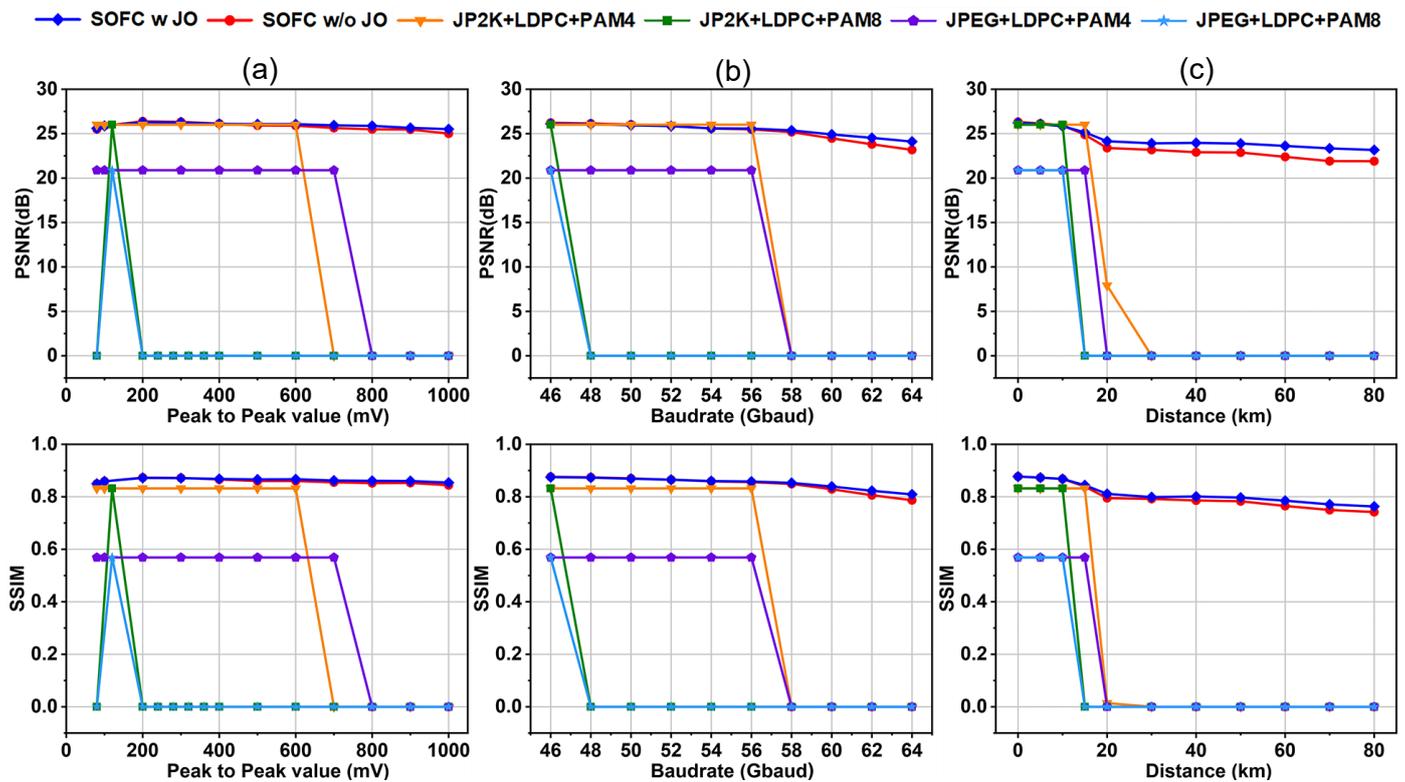

**Fig. 5** Transmission performance with respect to (a) peak-to-peak value, (b) baud rate, and (c) transmission distance.

## Conclusion

Transmission of semantic information rather than bits in optical fiber links could achieve great information compression and strong anti-noise ability. However, accurately extracting and restoring semantic information and adapting it to the optical physical channels are major challenges. Here, we propose and experimentally demonstrate a novel framework of SOFC system. The LA-net and DR-net were designed as semantic encoders for text and image transmissions, respectively. JO was performed to make the semantic information more adaptable to the transport in optical physical channels. The SOFC system showed significant advantages compared with the TOFC systems, especially in low ROP and high optical link impairment regimes.

The SOFC system achieved high-quality transmission in the low-SNR area by sacrificing some semantic information. This transmission based on semantics rather than bits makes it more robust against optical physical link impairments and avoid cliff-like performance degradation. Moreover, in the receiver of the SOFC system, the semantic decoder can be adjusted according to the optical physical channel state, enabling the interaction between information coding and optical physical layer, which is difficult to achieve in the TOFC systems. Thus, the SOFC system has the ability to break through the capacity limit of the existing optical fiber communication systems and achieve effective information transfer in extremely harsh channel environments. Furthermore, the SOFC system can be designed specially and integrated with a cross-domain solution according to different transmission tasks.

Such SOFC system is still in the initial research phase. The metrics used to evaluate the performance of semantic source reconstruction (*e.g.*, BLEU and SSIM) are still unable to clearly describe human perceptions. The formulation of a more intelligent and accurate metric is desired by researchers in the fields of communication, information theory, NLP, and CV. Different networks must be designed for different information sources separately, which means that unified semantic encoding is worth pursuing. Moreover, effective management of SOFC networks will also be a challenge because of the addition of semantic resources.

**Data and Materials Availability**

The data that support the findings of this study and custom codes are available from the corresponding author upon reasonable request.

**Acknowledgments**

This work was financially supported by the National Key R&D Program of China (No. 2021YFF0901700); the National Natural Science Foundation of China (No. 61821001, 61901045); and the Fund of State Key Laboratory of Information Photonics and Optical Communications, BUPT (No. IPOC2021ZT18).